\documentclass[11pt, a4]{article}
\usepackage{amssymb,amsmath,latexsym,graphicx, bm, mathrsfs}
\usepackage[hidelinks,citecolor=blue,urlcolor=blue,colorlinks=true,bookmarks=false,hypertexnames=true]{hyperref} 
\usepackage{natbib} %Bibliography
\setcitestyle{numbers,square} %Cite as numbers or author-year.
\usepackage{graphics,graphicx, cancel}
\usepackage[dvipsnames]{xcolor}
\usepackage{graphicx}
\graphicspath{ {./images/} }
\usepackage[british]{datetime2}
\usepackage[page,toc,titletoc,title]{appendix}
\usepackage{tocloft}
\usepackage{blindtext}
\usepackage{dsfont}
\usepackage{dirtytalk}

\usepackage{comment}
\usepackage{caption}
\usepackage{enumitem}
\usepackage{mathrsfs}
\usepackage{alphalph} % For better (a), (b), (c) labeling
\usepackage{changepage}
\newcounter{mainthm}

\newcounter{subthm}[mainthm]
\newtheorem{subtheoreminner}{Theorem}[subthm]

\begin{document}

\title{Isentropic processes for axisymmetric Black Holes
\vspace{-0.25cm} 
\author{Nitesh K. Dubey$ ^{a,b}$, Sanved Kolekar$^{a,b}$ \\ \vspace{-0.5cm} \\ 
\textit{$ ^a $Indian Institute of Astrophysics} \\ 
\textit{Block 2, 100 Feet Road, Koramangala,} 
\textit{Bengaluru 560034, India.} \\ 
\textit{$ ^b $Pondicherry University} \\ 
\textit{R.V.Nagar, Kalapet, Puducherry-605014, India}\\
\texttt{\small Email: \href{mailto:nitesh.dubey@iiap.res.in}{nitesh.dubey@iiap.res.in}, 
\href{mailto:sanved.kolekar@iiap.res.in}{sanved.kolekar@iiap.res.in}}}}

\maketitle

\abstract{We demonstrate that the isentropic absorption of a classical charged test particle is classically forbidden for all (3+1)-dimensional stationary, non-extremal, axisymmetric black holes in any diffeomorphism invariant theory of gravity. This result is derived purely from the near-horizon geometry and thermodynamic properties of the black hole spacetime, independent of the specific gravitational theory. We further consider the Kerr–Newman black hole in general relativity and analyse, using the quantum tunnelling approach, the conditions under which isentropic absorption may be allowed. Broader implications for the second law and extremality bounds are discussed.}

\pagebreak

\section{Introduction}
Beginning in the 1970s, several independent universal entropy bounds were proposed for arbitrary physical systems. Among the earliest was the Bekenstein bound, formulated using semiclassical arguments to suggest that the maximum entropy of an object with finite energy and spatial extent is bounded from above \cite{PhysRevD.9.3292, PhysRevD.7.2333, Bekenstein:1972tm, PhysRevD.23.287}. This result can also be interpreted as an upper limit on the information content of such a system \cite{PhysRevD.7.2333}. Since then, the Bekenstein bound has been explored and extended in a variety of directions. In particular, \cite{Hayden:2023ocd} examined how it constrains the communication capacity of quantum channels, while \cite{Bousso:2003cm} analysed its validity for bound states of interacting fields and showed that, once interactions are properly included, the bound remains intact. From a scattering perspective, \cite{Dvali:2020wqi} demonstrated that for a self-sustained quantum field object of size $R$, unitarity implies $S \leq 1/\alpha$, where $\alpha$ is the running coupling evaluated at the scale $1/R$. Black holes, in turn, are known to saturate this bound, with their entropy taking the form $S = 1/\alpha$, where $\alpha = G/R^{d-2}$ and $G$ is Newton’s constant in $d$ dimensions \cite{Dvali:2020wqi, Dvali:2011aa}. Additional derivations further reinforce the universality of this limit. For instance, \cite{Pesci:2009xh} showed from a holographic perspective that the Bekenstein bound holds for any strength of the gravitational coupling under certain assumptions, while a wide variety of generalised forms have also been proposed, including the causal entropy bound \cite{PhysRevLett.84.5695}, the Unruh–Wald bound \cite{PhysRevD.25.942, PhysRevD.27.2271}, generalized Bekenstein bounds \cite{Buoninfante:2020guu}, covariant entropy bounds \cite{PhysRevD.103.024002, Bousso1999}, and their generalized covariant versions \cite{Zhu:2023xbn, PhysRevD.62.084035}.

Alongside various formulations mentioned in the previous paragraph, different interpretations of the Bekenstein bound have also been explored. We recall some of them below. For instance, \cite{PhysRevD.102.106002} argued that it arises naturally from the Pauli exclusion principle if the fundamental degrees of freedom of a black hole are finite and fermionic. Related studies of such fermionic degrees of freedom are found in \cite{PhysRevX.5.041025, PhysRevD.94.106002}. Other works emphasise the implications of the bound for information dynamics: \cite{Acquaviva:2017krr} showed that the entropy bound governs the redistribution of degrees of freedom in generic systems and implies an average information loss due to entanglement between geometry and fields, while \cite{PhysRevD.49.1912} used it to constrain the information capacity of black hole remnants, arguing against remnants as a resolution of the information paradox. The interplay between entropy bounds and negative energy has also been highlighted. Since quantum fields admit regions of negative energy density provided they are compensated elsewhere, \cite{PhysRevLett.111.221601} showed that the Bekenstein bound requires such negative-energy regions to be localised near positive-energy ones. The consequences of negative energy densities for black hole dynamics have been further examined in \cite{HAJICEK19809, Balbinot:1986fx, Mann:1997jb}

All the considerations highlighted above are directly relevant for the black hole information loss paradox, which reveal tensions among fundamental principles: unitarity, local quantum field theory, general covariance, the identification of Bekenstein–Hawking entropy with Boltzmann entropy, and the existence of a semiclassical observer retrieving information from Hawking radiation cannot all simultaneously hold \cite{Ong:2016iwi, Mann:2025ojd, PhysRevD.48.3743, Raju:2020smc}. One possible resolution is that non-perturbative effects—though negligible in the early stages of evaporation—can accumulate and become significant over the black hole’s lifetime. Such effects may depend on spacetime dimensionality \cite{PhysRevD.102.026016}, memory burden phenomena \cite{PhysRevD.110.056029, PhysRevD.102.103523}, fluctuations in gravitational degrees of freedom \cite{PhysRevD.47.2446}, or critical instabilities \cite{PhysRevD.102.103523, PhysRevD.54.7490}. Although the non-perturbative mechanisms operate through distinct physical channels, they share a common capacity to accumulate effects gradually and ultimately perhaps dominate the entropy dynamics, thereby breaking the semiclassical predictions. Importantly, while the identification of Bekenstein–Hawking entropy with Boltzmann entropy is semiclassical \cite{PhysRevD.7.2333}, it has been argued in \cite{Mann:2025ojd} that non-perturbative contributions can break this equivalence and allow the Boltzmann entropy to exceed the Bekenstein–Hawking value. In particular, \cite{Mann:2025ojd} analysed the isentropic absorption of charged particles by a Reissner–Nordström black hole within the WKB approximation, suggesting that such a process may serve as a non-perturbative channel for exceeding entropy bounds. 

Motivated by the findings mentioned above, in the present article, we extend the analysis of the isentropic absorption of charged particles to stationary, axisymmetric black holes. We first demonstrate in Section \ref{sec:2} that, for non-extremal axisymmetric black holes, the absorption of a charged particle without changing the Bekenstein–Hawking entropy is classically forbidden: a classically forbidden region exists outside the horizon. The result of the existence of the classically forbidden region is rooted in the near-horizon geometry and thermodynamic laws--it holds across all diffeomorphism-invariant gravitational theories. We then turn, in Section \ref{sec:quantum}, to the quantum regime, where tunnelling across the barrier becomes possible. Using the saddle-point approximation, we compute tunnelling probabilities and show how they depend on both the particle’s and the black hole’s parameters.  Finally, Section \ref{sec:discussion}, delves into the broader significance of these processes: small, highly charged black holes emerge as more susceptible to isentropic absorption, potentially allowing entanglement (hence Boltzmann) entropy to grow—raising subtle questions regarding entropy bounds and information accumulation. We use units $\hbar = c = k_B = G = 1$ and adopt a mostly positive metric signature $(- + + +)$.

\section{Charged particle in stationary axisymmetric spacetime} \label{sec:2}
We consider the following general stationary axisymmetric black hole in 3+1 dimensions \cite{Chandrasekhar:1985kt, PhysRevLett.124.181101}:
\begin{equation} \label{eq:1}
    ds^2 = g_{tt}(r,\theta) dt^2 + 2 g_{t\phi}(r,\theta) dt d\phi + g_{\phi \phi}(r,\theta) d\phi^2 + g_{rr}(r,\theta) dr^2 + g_{\theta \theta}(r,\theta) d\theta^2.
\end{equation}
The metric functions are assumed to be smooth and continuous, at least $C^2$-differentiable, independent of the $t$ and $\phi$ coordinates, and thus adapted to the Killing symmetries of stationarity ($\partial_t$) and axisymmetry ($\partial_\phi$). The spacetime is assumed to be circular and asymptotically flat. Further,
the black hole spacetime is assumed to possess a Killing horizon at $r = r_+$ such that $g^{rr} (r_+,\theta) = 0$. The black hole is described by its mass $M$, charge $Q$, and angular momentum $J$. 

The trajectory of a particle of mass $ m $ and charge $q$ can be described by the following Lagrangian :
\begin{equation} \label{eq:2}
    \mathcal{L} = \frac{1}{2} m g_{\mu \nu}  \dot{x}^\mu \dot{x}^\nu + q A_\mu \dot{x}^\mu ,
\end{equation}
where $A_\mu = (-A_t(r,\theta),0,0,A_\phi(r,\theta))$ is the electromagnetic 4-vector potential arising from the stationary and axisymmetric black hole \cite{PhysRevD.87.124030}. Since $t$ and $ \phi$ coordinates are cyclic, the corresponding canonical momenta, say $E$ and  $L$ respectively, will be conserved. They are given by \footnote{At some places in literature, the canonical momentum conjugate to the cyclic coordinate $\phi$ is denoted by $L_z$. We write $L_z = L$.}
\begin{align} 
    E \equiv -  \frac{\partial\mathcal{L}}{\partial \dot{t}} = - m g_{tt} \dot{t} - m g_{t\phi} \dot{\phi} + q A_t \label{eq:3}\\
    L\equiv  \frac{\partial\mathcal{L}}{\partial \dot{\phi}} = m g_{\phi t} \dot{t} + m g_{\phi \phi} \dot{\phi} + q A_\phi \label{eq:4}
\end{align}
where dot denotes the derivative with respect to an affine parameter, say $\lambda$. Solving the above equation \eqref{eq:3}-\eqref{eq:4}, one gets the following equation of motion:
\begin{align}
    \dot{t} = \frac{E' g_{\phi \phi} + L' g_{t \phi}}{m D} \label{eq:5} \\
    \dot{\phi} = - \frac{E' g_{\phi t} + L' g_{t t}}{m D} \label{eq:6}
\end{align}
where, $D \equiv g_{t \phi}^2 - g_{tt}g_{\phi \phi}$, $E' \equiv E - qA_t$, $L' \equiv L - q A_\phi$. Now, for the classical test particle motion, one can rewrite Eq.\eqref{eq:1} as 
\begin{equation} \label{eq:7}
    - \epsilon = g_{tt}(r,\theta) \bigg( \frac{dt}{d\lambda}\bigg)^2 + 2 g_{t\phi}(r,\theta) \frac{dt}{d\lambda} \frac{d\phi}{d \lambda} + g_{\phi \phi}(r,\theta) \bigg( \frac{d\phi}{d\lambda}\bigg)^2 + g_{rr}(r,\theta) \bigg( \frac{dr}{d\lambda}\bigg)^2 + g_{\theta \theta}(r,\theta) \bigg( \frac{d\theta}{d\lambda}\bigg)^2,
\end{equation}
where $\epsilon$ is 1 for a massive particle. Taking $\epsilon$ to the right-hand side and using equations  \eqref{eq:5}-\eqref{eq:6}, one gets
\begin{equation} \label{eq:8}
    \bigg( \frac{dr}{d\lambda}\bigg)^2 + g^{rr} (r,\theta) g_{\theta \theta} (r,\theta) \bigg( \frac{d\theta}{d\lambda}\bigg)^2 + V_{eff}(r,\theta) = 0,
\end{equation}
where
\begin{equation} \label{eq:9}
     V_{eff}(r,\theta) = - \frac{ g^{rr} (E'^2g_{\phi\phi} + L'^2 g_{tt} + 2 g_{t\phi}L'E') }{m^2 D} + \epsilon  g^{rr}.
\end{equation}
The first and second terms in equation \eqref{eq:8} are positive, which implies that the third term, denoted by $V_{eff}$, must be negative for the expression to be valid for a classical test particle. Further, we stress that Eqs.\eqref{eq:8}-\eqref{eq:9} are valid for any diffeomorphism invariant theory of gravity that admits a black hole solution of the form described in \eqref{eq:1}.

\subsection{Isentropic absorption of the charged particle}
For an isentropic process, the second law of black hole mechanics leads to
\begin{equation} \label{eq:10}
    TdS = 0 = dM - \Phi_edQ - \Omega dJ.
\end{equation}
The change in mass $dM$, charge $dQ$, and angular momentum $dJ$ of the black hole, when the charged particle enters the horizon, is given by $E$, $q$, and $L$, respectively. The spacetime under consideration is stationary, so the event horizon will be a Killing horizon \cite{Date2001, Carroll_2019}. One can consider the Killing horizon to be generated by the Killing vector 
\begin{equation} \label{eq:13}
    \chi^\mu = \partial_t + \Omega_H \partial_\phi .
\end{equation}
Here, $\Omega_H$ denotes the angular velocity evaluated at the outer horizon of the black hole, which is interpreted as the angular velocity of the black hole itself.  The electric potential on the horizon is defined by $ \Phi_{e} \equiv -\,\chi^{\mu} A_{\mu}\big|_{r_{+}} $ \cite{Caldarelli:1999xj}.
Using our convention in the Lorentz gauge for the gauge field, $A_{\mu}=(-A_{t},0,0,A_{\phi})$,
the contraction gives \( \Phi_{e}=A_{t}-\Omega_H A_{\phi},\)
so that the change in electromagnetic potential energy of the black hole due to absorption of a particle of charge $q$ is $q\Phi_{e}=qA_{t}-\Omega_H qA_{\phi}$. Therefore, the above equation \eqref{eq:10} gives 
\begin{equation} \label{eq:11}
    E- qA_t + q \Omega_H A_\phi - \Omega_H L = 0,
\end{equation}
which is $E' = \Omega_H L'$, at the event horizon. Therefore, the effective potential at the horizon can be written as
\begin{equation} \label{eq:12}
     V_{eff}(r=r_+,\theta) = - \frac{ g^{rr} L'^2 (\Omega_H^2g_{\phi\phi} +  g_{tt} + 2 g_{t\phi} \Omega_H) }{m^2 D}\bigg|_{r=r+} + \epsilon  g^{rr} \bigg|_{r=r+}
\end{equation}
The above expression for the effective potential at the horizon in terms of the Killing vector $\chi^\mu$ can be rewritten as
\begin{equation} \label{eq:14}
     V_{eff}(r=r_+,\theta) = - \frac{ g^{rr} L'^2 \chi^\mu\chi_\mu }{m^2 D} \bigg|_{r=r+} +  \epsilon  g^{rr} \bigg|_{r=r+}.
\end{equation}

\begin{figure}
    \centering
    \includegraphics[width=.92\textwidth]{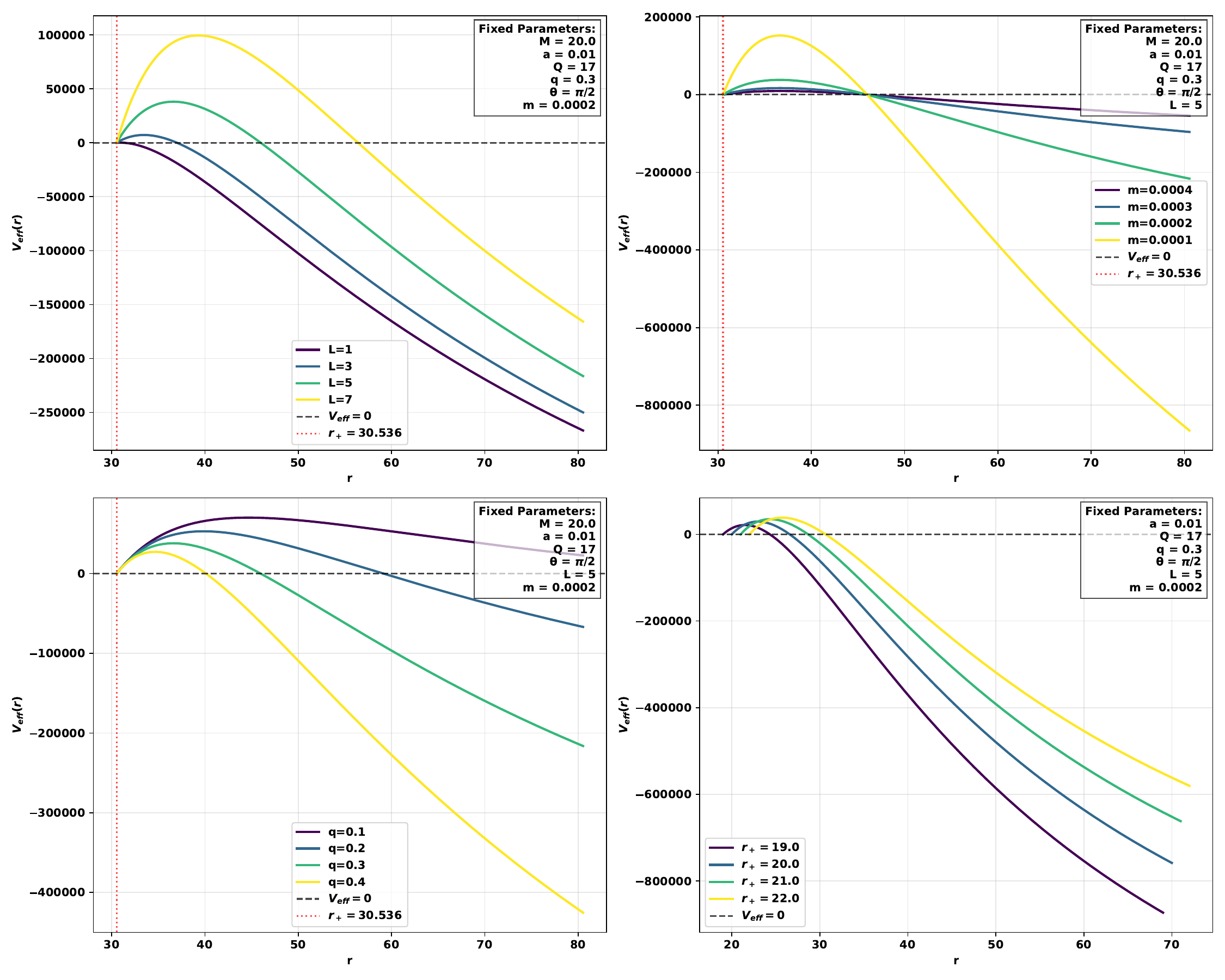}
     \captionsetup{margin=1cm, font=small}
    \caption{The above plots show the effective potential experienced by a charged particle for isentropic absorption by a Kerr-Newman black hole in general relativity. The expression for the effective potential is given in Eq.\eqref{eq:9}, with $E'(r) \equiv E - qA_t(r)= qA_t(r_+) + \Omega(L - qA_\phi(r_+)) -qA_t(r)$ for the isentropic absorption. The unit of $\theta$ is taken to be radians, so the plots are near the equator.}
    \label{fig:effective potential}
\end{figure}
The effective potential given above in Eq.\eqref{eq:14} is independent of the sign of $L'$. One can use the property of axisymmetry of the spacetime and the zeroth law of black hole mechanics to show that $\chi^\mu \chi_\mu$, $g^{rr}$, and $D$ all vanish linearly at the Killing horizon of a stationary axisymmetric non-extremal black hole \cite{PhysRevD.70.024009}, i.e., as follows:
\begin{align} \label{nearh}
\chi^\mu \chi_\mu \approx - 2 \kappa (r - r_+) + \mathcal{O} ((r - r_+)^2) ;D \approx 2 \kappa (r - r_+) g_{\phi \phi}  + \mathcal{O} ((r - r_+)^2) ; g^{rr} \propto  (r - r_+).
\end{align}
Here, $\kappa$ is the surface gravity which is constant over the Killing horizon from the zeroth law of black hole mechanics \footnote{One can refer to \cite{ArderucioCosta:2019otn} for a proof of the zeroth law of black hole mechanics that does not depend on any assumptions about the dynamics of the theory, but only assumes the existence of a Killing horizon with a nonzero $\kappa$.}. Hence, the above effective potential $V_{eff}(r_+,\theta)$ vanishes at the horizon for an isentropic absorption of the test charged particle. Taking the derivative of Eq.\eqref{eq:9} with respect to $r$ one gets after some manipulation:
\begin{equation} \label{eq:15}
    \frac{dV_{eff}}{dr} = g'^{rr} g_{rr} V_{eff} - \frac{D'}{D} (V_{eff} - g^{rr} \epsilon) - \frac{g^{rr}}{m^2D} (E'^2 g'_{\phi\phi} + L'^2 g'_{tt}+ 2 L'E'g'_{t\phi}),
\end{equation}
where prime on $D$ and the metric components denote the derivative with respect to $r$. Evaluating the above expression on the horizon for an isentropic process with $E' = \Omega L'$, one gets,
\begin{equation} \label{eq:16}
    \frac{dV_{eff}}{dr} \bigg|_{r=r_+} = \frac{\epsilon D' g^{rr}}{D} \bigg|_{r=r_+} + \frac{L'^2g^{rr}}{m^2 D} \bigg[  \frac{D'\chi^\mu\chi_\mu}{D} - \frac{g'^{rr}\chi^\mu\chi_\mu}{g^{rr}} - (\chi^\mu\chi_\mu)_{,r} \bigg]\bigg|_{r=r_+}
\end{equation}
Now, using L'Hospital rule for taking limit near horizon and the property that $D$, $\chi^\mu\chi_\mu $, and $g^{rr}$ all go to zero near horizon linearly as shown in Eq.\eqref{nearh}, for the nonextremal case, one gets
\begin{equation} \label{eq:17}
     \frac{dV_{eff}}{dr} \bigg|_{r=r_+}  = g'^{rr} \bigg[  \epsilon - \frac{ L'^2  (\chi^\mu\chi_\mu)_{,r}}{m^2 D'}  \bigg]_{r=r_+}  = g'^{rr} \bigg[  \epsilon + \frac{ L'^2 }{ m^2 g_{\phi\phi}}  \bigg]_{r=r_+}  >0,
\end{equation}
since $g'^{rr} > 0$ and $g_{\phi \phi}>0$ at the horizon. Since $V_{eff}(r_+,\theta) = 0$ and $dV_{eff}(r_+,\theta)/dr >0$, we necessarily have $V_{eff} > 0$ outside the horizon and hence, classically, it is not possible to absorb the charged particle isentropically.

At a large distance from the black hole, the electromagnetic 4-vector potential vanishes, i.e,   $A_t(r\rightarrow \infty,\theta) \rightarrow 0, A_\phi(r\rightarrow \infty,\theta) \rightarrow 0 $. Therefore, $E'\rightarrow E, L'\rightarrow L$ at spatial infinity. Further, $g^{rr} \rightarrow 1$, $g_{tt} \rightarrow -1, g_{t\phi }\rightarrow 0, g_{\phi\phi} \rightarrow \infty$ at spatial infinity. So, the effective potential shown in Eq.\eqref{eq:9} takes the following form:
\begin{equation} \label{assymptotic}
    V_{eff}(r \rightarrow \infty, \theta) \rightarrow - \frac{E^2}{m^2} + \epsilon.
\end{equation}
The above asymptotic expression for the effective potential is consistent with that expected for a flat spacetime. For a particle at rest at infinity, Eq.~\eqref{assymptotic} gives \( V_{\text{eff}}(r \rightarrow \infty, \theta) \rightarrow 0 \), while it becomes negative for a moving particle. For the isentropic absorption, the conserved energy $E = \Omega L + q A_t(r=r+,\theta) - \Omega q A_\phi(r=r+,\theta) >m^2$, which is possible for an appropriate choice of parameters $\{q,m,L,Q,M,a \}$. So, the effective potential at infinity is negative for this set $\{q,m,L,Q,M,a \}$, which implies that there also exists a classically allowed region for $r > r_2$ for some $r_2 > r_+$ apart from the classically forbidden region for $r_+ < r < r_2$.

\subsection{Kerr-Newman black holes in general relativity} \label{sec: kerr-newman}
So far, our discussion regarding the isentropic absorption of a charged particle by a stationary axisymmetric nonextremal black hole was quite general. For the sake of definiteness, we now restrict to the non-extremal Kerr-Newman black hole in general relativity \footnote{Since the surface gravity, $\kappa$, is zero in the extremal case, the discussion in the previous section may not be true for the extremal Kerr-Newmann black hole.}.  In general relativity, the Kerr-Newman black hole in 3+1 dimensions, in Boyer–Lindquist coordinates, is described by the metric \cite{Chandrasekhar:1985kt, PhysRevD.87.124030}
\begin{align} \label{metric}
ds^2 = &- \frac{\Delta - a^2 \sin^2\theta}{\Sigma} \, dt^2 
- \frac{2a \sin^2\theta (r^2 + a^2 - \Delta)}{\Sigma} \, dt \, d\phi 
+ \frac{(r^2 + a^2)^2 - \Delta a^2 \sin^2\theta}{\Sigma} \sin^2\theta \, d\phi^2 \nonumber \\
&+ \frac{\Sigma}{\Delta} \, dr^2 
+ \Sigma \, d\theta^2,
\end{align}
with
\begin{align}
\Sigma(r,\theta) &= r^2 + a^2 \cos^2\theta, \label{eq:19}\\
\Delta(r) &= r^2 - 2Mr + a^2 + Q^2  \label{eq:20}.
\end{align}
Here, \( a \) is the specific angular momentum, \( M > 0 \) is the ADM mass, and \( Q \) is the electric charge of the black hole. Assuming a vanishing magnetic charge, the electromagnetic 4-vector potential of the black hole is given by $A_\mu= (-Q r/\Sigma, 0,0, Q a r \sin^2\theta/\Sigma)$. So, the sign convention is such that a positive value of $Q$ means a positively charged black hole.
By substituting the metric in Eq.\eqref{metric} into Eqs.\eqref{eq:12} and \eqref{eq:15}, one obtains, respectively, for the isentropic absorption of a charged particle,
\begin{align}
    V_{eff}\bigg|_{r=r_+} &= 0 \label{eq:21} \\
    \frac{dV_{eff}}{dr}\bigg|_{r=r_+} &= \frac{2(Mr_+ -a^2-Q^2)}{r_+^3} \bigg[ 1 + \frac{L'^2 (r_+^2+a^2 \cos^2\theta)}{(r_+^2 + a^2)^2}  \bigg] >0. \label{eq:22}
\end{align}
The positivity of $dV_{eff}/dr$ at $r = r_+$ in Eq.~\eqref{eq:22} follows from the fact that the term inside the square brackets is strictly positive, while the prefactor is also positive, since $ Mr_+ = M^2 + M \sqrt{M^2 - a^2 - Q^2}, $ and the non-extremality condition requires $M^2 > a^2 + Q^2$. Therefore, the isentropic absorption of the charged test particle is not possible classically.
\begin{figure}
    \centering
    \includegraphics[width=.92\textwidth]{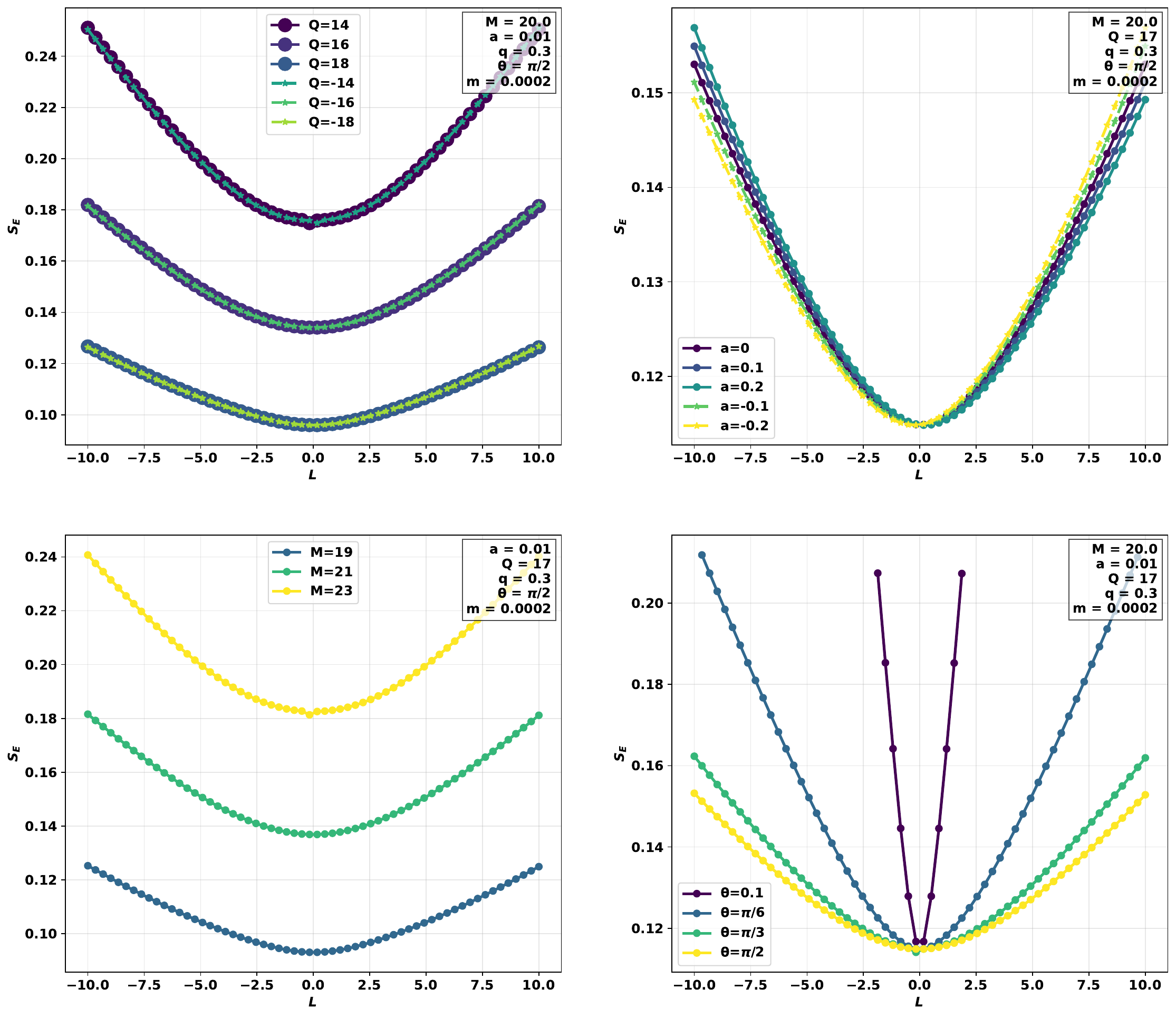}
     \captionsetup{margin=1cm, font=small}
    \caption{The above plots show the Euclidean action experienced by a charged particle for isentropic absorption by a Kerr-Newman black hole in general relativity. The turning points exist only for certain values of parameters. The plots used the parameter values for which the turning point exists.}
    \label{fig: L vs effective action}
\end{figure}

\begin{figure}
    \centering
    \includegraphics[width=.92\textwidth]{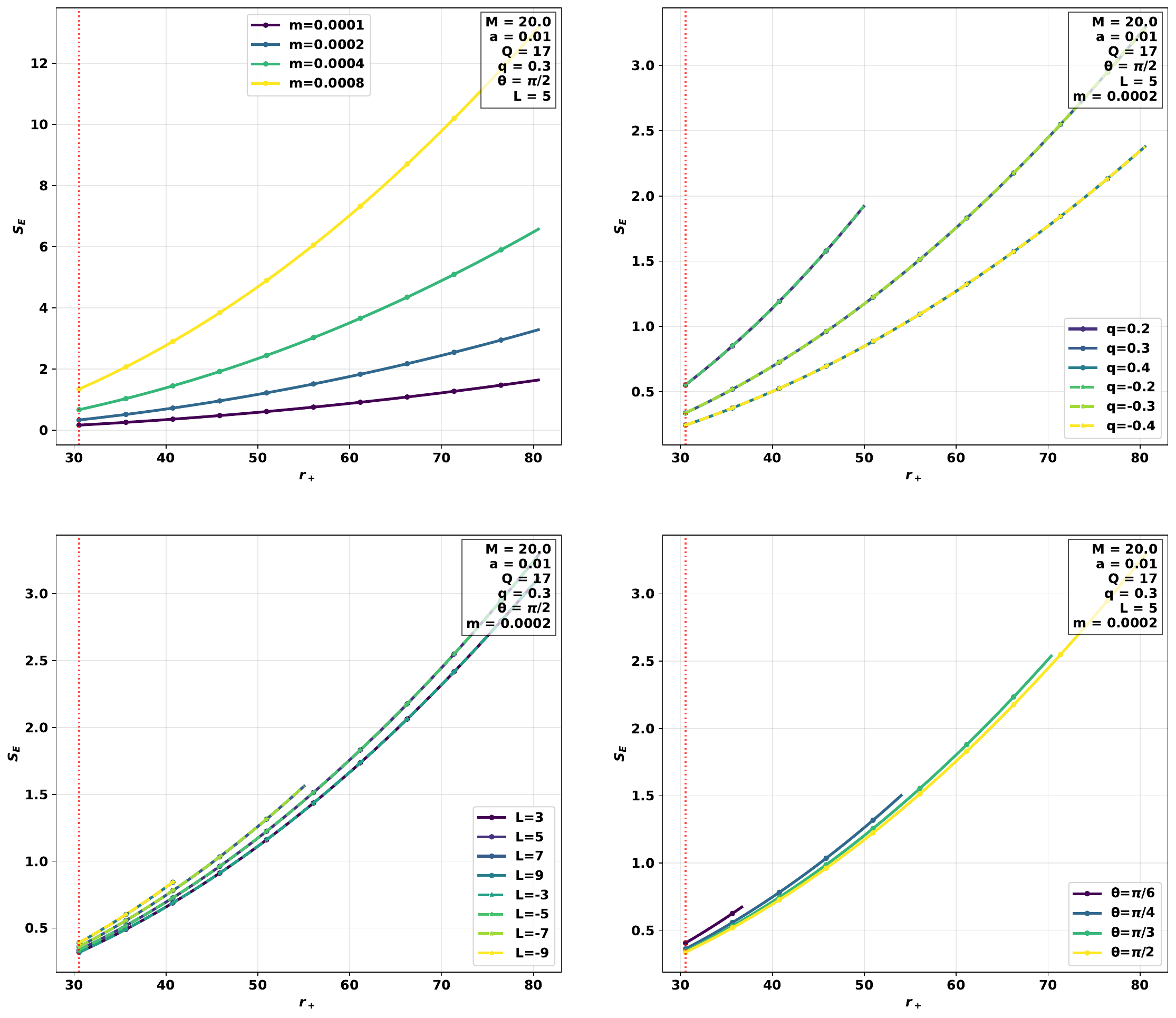}
     \captionsetup{margin=1cm, font=small}
    \caption{The above plots show the Euclidean action experienced by a charged particle for isentropic absorption vs the radius of the outer horizon of a Kerr-Newman black hole in general relativity. The turning points exist only for certain values of parameters. The plots used the parameter values for which the turning points exist.}
    \label{fig: horizon vs effective action}
\end{figure}

\section{Quantum tunneling} \label{sec:quantum}
We saw in the previous section that the effective potential vanishes at the event horizon and that its derivative becomes positive for the isentropic absorption. We plot the effective potential in Fig.[\ref{fig:effective potential}], from which it can be seen that for certain values of charge, mass, and angular momentum, the effective potential can become negative again at some distance from the event horizon. The region between the event horizon at $r=r_+$ and the point outside, say $r=r_2>r_+$, where the effective potential vanishes again, is classically forbidden. We see from Fig.~[\ref{fig:effective potential}] that increasing the angular momentum of the particle increases both the width of the classically forbidden region and the height of $V_{eff}$. However, the increase in charge reduces both. Further, we see that the change of mass of the test particle does not change the width of the classically forbidden region, but the height of the potential barrier experienced changes. So, the massive charged test particles with low angular momentum experience a small, but nonzero, potential for the isentropic absorption.

Feynman's path integral formulation expresses the transition amplitude for a radial degree of freedom $r(\lambda)$, evolving in an effective potential $V_{eff}$, as
\begin{equation}
K \; \propto \; \int \mathcal D r \; \exp\!\left(\tfrac{i}{\hbar} S\right),
\end{equation}
where $S$ is the action. The quantum dynamics is found by summing over all admissible trajectories $r(\lambda)$, each weighted by the phase factor $\exp(iS/\hbar)$ determined by the action. In the classically forbidden region, where $V_{eff} > 0$, the path integral is dominated by the steepest-descent contour obtained via a Wick rotation to imaginary time, $\lambda \to -i\lambda_E$. Under this transformation, the action becomes Euclidean, $S \to iS_E$, and the amplitude reduces to
\begin{equation}
K \; \sim\; e^{-S_E/\hbar}, 
\qquad 
S_E = \int d\lambda_E \, L_E\,,
\end{equation}
where $L_E$ is the Euclideanized Lagrangian. The dominant contribution comes from the Euclidean classical trajectory---the solution of the Euler--Lagrange equations in imaginary time---connecting the turning points of the potential barrier. Since the amplitude acquires a factor $e^{-S_E/\hbar}$, the tunnelling probability is obtained by taking the modulus squared of this amplitude,
\begin{equation}
\Gamma \;\propto\; |K|^2 \;\sim\; e^{-2S_E/\hbar}.
\end{equation}
So, setting $\hbar = 1$, the probability through the forbidden region 
$[r_{2}, r_{+}]$ can be expressed as \cite{Mann:2025ojd}
\begin{equation} \label{eq:23}
    \Gamma \;\approx\; e^{-2S_E}.
\end{equation}
Here, the Euclidean action $S_E$ is given by
\begin{align}
    S_E &= \int _{r_2} ^{r_+} L_E d\lambda_E \\
    & = \int _{r_2} ^{r_+} \frac{1}{dr/d\lambda_E}d r \\
    & = - \int _{r_2} ^{r_+} \frac{1}{\sqrt{V_{eff}(r)}} dr \\  & = \int _{r_+} ^{r_2} \frac{1}{\sqrt{V_{eff}(r)}} dr  \label{eq:28}.
\end{align}
Here, we have assumed $d\theta / d\lambda = 0$ for simplicity and have used Eq.~\eqref{eq:8}. We choose the negative root in $dr/d\lambda_E = \pm \sqrt{V_{eff}}$ because the particle is being absorbed by the black hole. It is difficult to evaluate the integral in Eq.~\eqref{eq:28} analytically, so we compute it numerically and plot the corresponding Euclidean action for a real value of turning points $r_+$ and $r_2$. The tunnelling probability in Eq.\eqref{eq:23} decreases (increases) as the Euclidean action increases (decreases), so Fig.[\ref{fig: L vs effective action}]-[\ref{fig: horizon vs effective action}] effectively describe the tunnelling probability in the approximation discussed above.  

One can notice from Fig.[\ref{fig: L vs effective action}] that the Euclidean action increases with an increase in the angular momentum of the particle for all combinations of parameters. Therefore, the tunnelling probability for low angular momentum particles is large in comparison with the higher angular momentum particles. One can understand it as the effective potential, as well as the width of the classically forbidden region, increasing with an increase in the angular momentum of the test particle. Furthermore, the increase in the magnitude of charge $Q$ of the black hole reduces the Euclidean action. However, the increase in mass of the black hole increases the Euclidean action. The increase of the specific angular momentum $a$ of the black hole decreases the effective action for the positive $L$ particles, but increases the effective action for the negative $L$ particles. Therefore, a low mass, high specific angular momentum, and large charge black hole has a higher tunnelling probability for a test charged particle. The Euclidean action is smaller for a larger polar angle. Therefore, the isentropic absorption is more probable near the equator of the black hole.

In Fig.[\ref{fig: horizon vs effective action}], we see that the increase in the size of the black hole, determined by its outer horizon $r_+$, increases the Euclidean action for all combinations of parameters. Therefore, smaller black holes have a greater tunnelling probability than larger black holes. This suggests that the smaller black holes are \textit{relatively less classical} as they can absorb charged particles isentropically and are prone to violate the entropy bound more efficiently than a large black hole. Furthermore, we notice that increasing the mass and the magnitude of angular momentum of the test particle increases the Euclidean action for a fixed size $r_+$ of the black hole, whereas an increase in the magnitude of the charge of the test particles decreases the Euclidean action. Therefore, lighter particles with small angular momentum, but large charge, have a large tunnelling probability. Further, the tunnelling probability is larger near the equator for a fixed $r_+$.

\section{Discussion} \label{sec:discussion}
We investigated whether a stationary axisymmetric black hole can absorb a classical test charged particle without a change in the black hole's entropy, and found that this process is classically forbidden in any diffeomorphism-invariant theory of gravity. We found that the conditions required to keep the Bekenstein–Hawking entropy unchanged lead to a classically inaccessible region outside the Killing horizon. Further, we analysed a Kerr-Newman black hole as a specific example. A test charged particle outside a Kerr–Newman black hole experiences an effective potential for isentropic absorption, which remains positive between the outer horizon and a certain distance away from it. So, classically, this region outside the black hole is forbidden. However, for a certain range of mass, charge, and angular momentum of a test charged particle, it is quantum mechanically possible to tunnel through this classically forbidden region.

Our results show that the tunnelling probability increases with a decrease in the mass of either the test particle or the black hole. However, the tunnelling probability increases with an increase in the charge of either the test particle or the black hole. Furthermore, we also find that the tunnelling probability is large for a small black hole in comparison to the larger black holes. These observations suggest that the low mass and high charge black holes with a smaller radius of the outer horizon are more prone to isentropic absorption of classical test charged particles. Moreover, if a charged particle that is entangled with degrees of freedom outside the black hole is isentropically thrown into the black hole, the entanglement between the inside and outside of the black hole increases without any change in the areal entropy. As emphasised in \cite{Mann:2025ojd}, this raises important questions about whether the black hole can accumulate entanglement entropy isentropically over time. Since the Boltzmann entropy accounts for the total number of accessible microstates and is typically greater than the entanglement entropy \cite{Mann:2025ojd, PhysRevD.102.125025, PhysRevA.101.052101, PhysRevE.87.042135}, could it perhaps happen through such a process that the Boltzmann entropy becomes greater than the Bekenstein-Hawking entropy? 

From the perspective of information theory, the existence of such channels suggests that the information content associated with black holes may not always be constrained in the strict semiclassical sense. This can be contrasted with earlier work linking entropy bounds to information capacity \cite{Hayden:2023ocd, PhysRevD.49.1912} and to the redistribution of fundamental degrees of freedom \cite{Acquaviva:2017krr}. Taken together, our results reinforce the view that the entropy-bound relation remains robust at the semiclassical level but may admit controlled, non-perturbative violations in black hole systems. Such violations are not pathological but instead highlight the limitations of semiclassical reasoning when extended over a black hole’s lifetime. They may also play a role in addressing the long-standing black hole information paradox \cite{Raju:2020smc, Ong:2016iwi}.

We discussed the isentropic absorption of charged particles by axisymmetric, non-extremal black holes in (3+1) dimensions. How our results change for extremal black holes remains a topic for future investigation, as this scenario may reveal new features of black hole thermodynamics and quantum gravity. Extending this analysis to rotating or charged black holes in higher dimensions may also uncover rich dynamics that are inaccessible in simpler (3+1)-dimensional models. We employed the saddle-point approximation of path integrals to evaluate the tunnelling probability, which provides a framework to study the leading-order contribution to the particle absorption beyond classical absorption. One could further explore the effects of loop corrections, as well as backreaction effects, on the tunnelling probabilities, potentially leading to refined predictions for black hole evaporation rates. Another direction is to investigate whether these non-perturbative channels leave observable imprints in late-stage Hawking radiation or in quantum-gravitational corrections to entropy bounds, which could provide novel tests for our understanding of information loss in black holes. 

\bibliography{reference}

\end{document}